\documentclass{aastex}
\usepackage{epsfig}
\usepackage{emulateapj5}

\makeatletter

\newenvironment{inlinefigure}{%
\def\@captype{figure}%
\noindent\begin{minipage}{0.999\linewidth}\begin{center}}
{\end{center}\end{minipage}\smallskip}
\makeatother

\def\spose#1{\hbox to 0pt{#1\hss}}
\def\lax{$\mathrel{\spose{\lower 3pt\hbox{$\mathchar"218$}}
     \raise 2.0pt\hbox{$\mathchar"13C$}}$}
\def\gax{$\mathrel{\spose{\lower 3pt\hbox{$\mathchar"218$}}
     \raise 2.0pt\hbox{$\mathchar"13E$}}$}
\newcommand{\chandra}{{\it Chandra}}
\newcommand{\xmm}{{\it XMM-Newton}}
\newcommand{\lum}{\thinspace\hbox{$\hbox{erg}\thinspace\hbox{s}^{-1}$}}

\begin{document}

\title{\chandra\ Studies of the X-ray Point Source Luminosity
Functions of M31}
\author{Albert K.H.~Kong} 
\affil{Harvard-Smithsonian Center for Astrophysics, 60
Garden Street, Cambridge, MA 02138}
\author{Rosanne Di\,Stefano}  
\affil{Harvard-Smithsonian Center for Astrophysics, 60
Garden Street, Cambridge, MA 02138}
\affil{Department of Physics and Astronomy, Tufts
University, Medford, MA 02155}
\author{Michael R.~Garcia}
\affil{Harvard-Smithsonian Center for Astrophysics, 60
Garden Street, Cambridge, MA 02138}
\author{Jochen Greiner}
\affil{Max-Plank-Institut fur extraterrestrische Physik, Postfach 1603, 
85740 Garching, Germany}

\begin{abstract}
Three different M31 disk fields, spanning a range of stellar
populations, were observed by
\chandra. We report the X-ray point source luminosity function
(LF) of each region, and the LF of M31's globular clusters, 
and compare these with each
other and with the LF of the galaxy's bulge.   
To interpret the results we also consider
tracers of the stellar population, such as OB associations 
and supernova remnants. We find differences in the LFs 
among the fields, but cannot definitively relate them to the
stellar content of the fields. We find that stellar 
population information, average and maximum source luminosities,
X-ray source densities, and slopes of the LF are useful in combination.  

\end{abstract}


\section{Introduction}

The luminosity function (LF) of point X-ray sources can provide useful
information about galactic stellar populations. \chandra's good
angular resolution allows detailed X-ray source
population studies of
external galaxies. 
Because M31 is close ($780$ kpc; 
Stanek \& Garnavich 1998; Macri et al. 2001), we can probe 
its LF to luminosities as low as
a few times $10^{35}$ \lum\ with $\sim 15-45$
ksec {\it Chandra} exposures. M31 is therefore an excellent
extragalactic spiral
galaxy in which to study X-ray source
populations with a wide range of luminosities.
Previous studies of M31's LF ({\it Einstein} [Trinchieri \&
Fabbiano 1991], {\it ROSAT} [Primini, Forman, \& Jones 1993; Supper et
al. 1997], \xmm\ [Shirey et al. 2001] and \chandra\ [Kong et
al. 2002a; Kaaret 2002]) have concentrated on the central
region ($< 30'$). In particular, Kong et al. (2002a) found that the
shape of LF varies with radius even in the central $17'\times17'$ region.
More recently, Trudolyubov et al. (2002)
used \xmm\ to study the LF of the northern disk, 
and found
it to differ from that of the central region,
suggesting a different source population.

In this paper, we present the LFs of three M31 disk regions 
as observed by \chandra.  We compare them to LFs of the
central region (Kong et al. 2002a) and of globular clusters 
(GCs; Di\,Stefano et al.\ 2002a).
We find that the  
slopes of the LFs are of limited use in relating
X-ray source populations to stellar populations. 
We 
therefore include in our study other quantities derived from the X-ray data
(e.g., source densities, and average and maximum luminosities) as
well as optical
tracers of the stellar population.

\section{Observations and Data Reduction}

Figure 1 shows the
fields-of-view of the \chandra\ ACIS-S\,\footnote{Details about ACIS can
be found at http://asc.harvard.edu/udocs/docs/POG/MPOG/node11.html}
observations of M31.
Three disk regions of M31 were chosen to span a wide
range of M31 stellar populations; each was observed
$3$ times ($\sim 15$ ksec for each ACIS-S observation)
at intervals of $3-4$ months during 2000-2001 (PI: Di\,Stefano). A
detailed description of the observations will be presented in a
companion paper.

Field 1, centered at R.A.=00h38m37s, Dec.=+40$^{\circ}$17$^m$41$^s$,
is farthest from the galaxy
center. Field 3 is closer to the galaxy center, but the aim point
(R.A.=00h46m17s, Dec.=+41$^{\circ}$41$^m$07$^s$) is offset
from the long axis of M31's disk. Those parts of Field 3  
closest to the central axis encompass an arc of M31's
10 kpc star-forming ring (e.g., Haas et al. 1998). Because the  portion of Field 3
away from the ring contains an apparently distinct, older, stellar
population, we have divided Field 3 into 2 distinct parts:
Field 3A 
includes the star-formation ring 
(X-ray sources in Field 3A are shown in red), 
while 
Field 3B is the largest portion of the field and does not
include the ring.
Field 2 is close to the center of the galaxy (R.A.=00h41m53s,
Dec.=+41$^{\circ}$00$^m$45$^s$). Part of it 
overlaps the bulge, and the central $17'\times17'$ region that
has been studied with \chandra\ ACIS-I (Kong et al. 2002a; red polygon
in Figure 1). Because 
the part of this region closest to
the nucleus is far off axis ($> 15'$) in a crowded field,
we consider it separately, calling it Field 2B (white dots in Figure
1), while the larger part of Field 2, containing the aim point, is
Field 2A. 
There are 12 GC X-ray sources in Fields 1, 2A and 3A+3B.

\begin{figure*}[htb]
\centerline{
\psfig{file=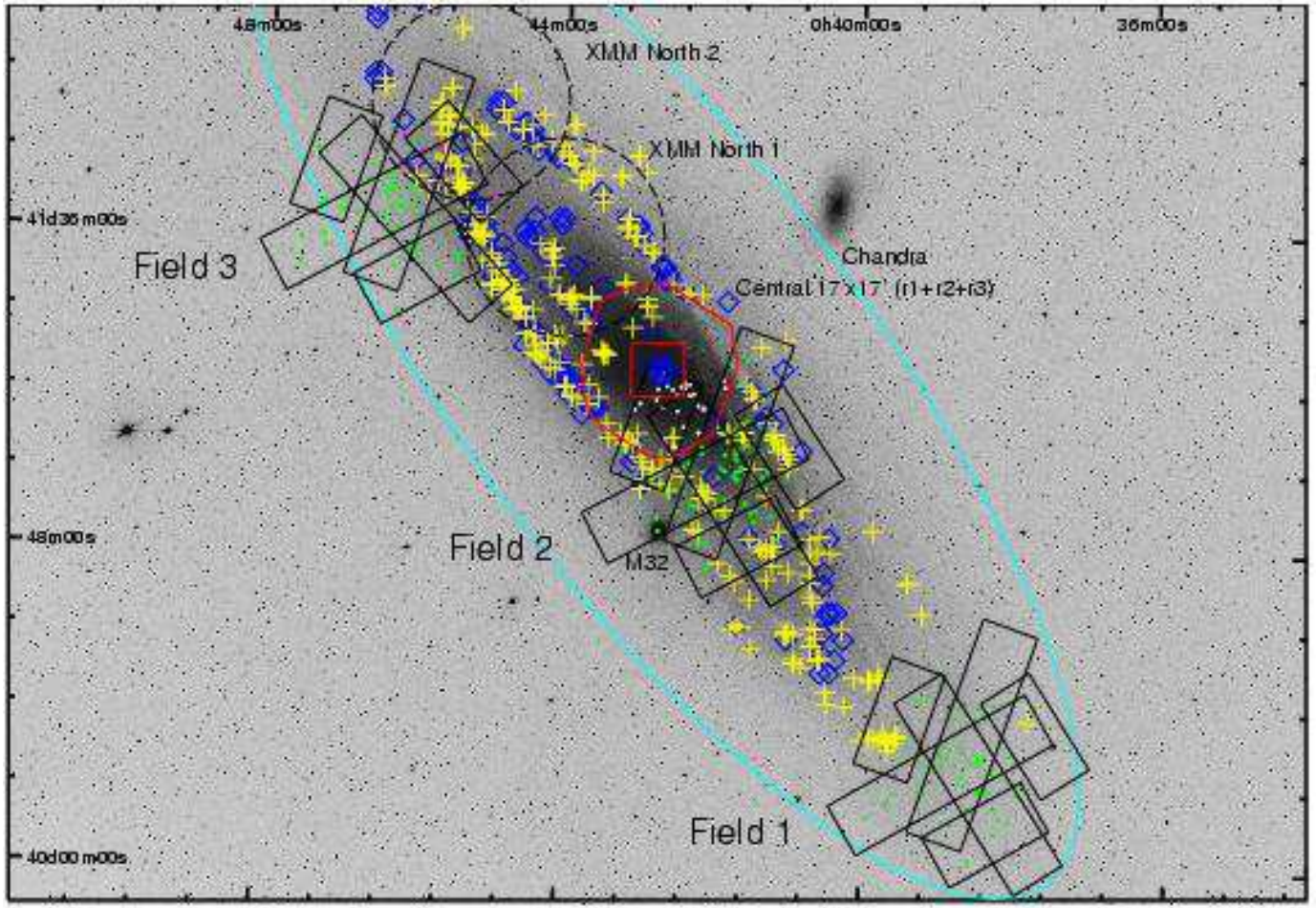,height=5in}
}
\caption{The regions observed by \chandra\ ACIS-S and the detected
X-ray sources (green dots for Fields 1, 2A and 3B; white dots for 2B;
red dots for 3A) overlaid on an optical Digital Sky Survey image of
M31. The fields-of-view of the two
\xmm\ observations (dashed circles) and \chandra\ ACIS-I observations
(red polygon)
of the central region (Kong et al. 2002) are also shown; the central
red square inside the polygon is the central $8'\times8'$ region  
(r1+r2 in Kong et al. 2002). The asymmetric shape of the ACIS-I region is
due
to combining observations with different roll angles (see Kong et al.
2002). Also shown in the
figures are the optical position of supernova remnants (yellow crosses)
and OB associations (blue diamonds). The ellipse show the $D_{25}$
size of M31.
The location of M32 is marked. North is up, and east is to
the left.}
\end{figure*}

For each observation, we examined the background and
rejected all high background intervals. Only events with photon
energies in the range of 0.3--7 keV were included in our analysis. The
three observations in each
field were merged. To detect sources we used
CIAO task {\sc
wavdetect} (Freeman et al. 2002). Source count rates were determined
via aperture photometry and were corrected for effective 
exposure and vignetting. 
The radius of the aperture was varied with the
average off-axis angle to match the 90\% encircled energy
function. 
Background was extracted from an annulus centered on each
source. Every extraction region was examined carefully in the images,
and in some cases, we had to modify the extraction region to avoid
nearby sources. 
Sources clearly associated with M32 are excluded from this analysis. 
Obvious foreground stars (see Di\,Stefano et al. 2002b for details) were excluded. The 
total number of sources detected in
Field 1, 2 and 3 is 53, 99 and 65 respectively.
Detailed source lists and source properties of all the
sources in our fields will be
presented in forthcoming papers.

\section{X-ray Luminosity Functions of the X-ray Sources}

The background subtracted count rates were converted to luminosities by
assuming an absorbed power-law energy spectrum with photon index 1.7,
column density $N_H=10^{21}$ cm$^{-2}$ and a distance of 780 kpc. 
This model is consistent 
with previous \xmm\ (Shirey et al. 2001) and \chandra\ (Kong et
al. 2002a) observations, in which the spectra of many bright point
sources could be well fit by similar models. Although this
type of spectral model does not apply to certain classes of sources,  
such as supersoft sources (Di\,Stefano \& Kong, in preparation) and
supernova remnants (e.g., Kong et al. 2002b), the number of such 
sources in our data sets is small. In general,
the luminosities derived by choosing different  
spectral models (e.g., thermal bremsstrahlung) would differ from the
ones we derive by less than a factor of $\sim 2$, except for supersoft sources. Considering 
only power-law spectral models and varying 
the photon index from $1.2$ to $2.5$, and sampling values of $N_H$
from $7\times10^{20}$ 
cm$^{-2}$ to $5\times10^{21}$ cm$^{-2}$, we find that our luminosity
results can be expressed as $L^{+1.2\, L}_{-0.2\, L},$ where $L$ is the
luminosity in 0.3--7 keV, and the superscript (subscript) is the
maximum (minimum) luminosity associated with any of the models
we considered.

Figure 2 shows the cumulative LFs for X-ray
sources (excluding GCs) in the three fields. 
Although the detection
limit of our observations is about $10^{35}$ erg s$^{-1}$, the LFs
flatten below $10^{36}$ erg s$^{-1}$, because the exposure
times vary across the images. Following Kong et al. (2002a), we computed histograms
of the number of detected sources against 
signal-to-noise ratio (S/N) to examine the completeness limit; the histograms peak 
at S/N$\sim 7$, corresponding to $\sim 10^{36}$ erg s$^{-1}$, and fall off below this. 
Hence, the LFs are  complete down
to $10^{36}$ erg s$^{-1}$. Also plotted in Figure 2 are the
LFs of X-ray emitting globular clusters (GCs) and of the bulge region (Kong et
al. 2002a).

We used a maximum likelihood method
(e.g., Crawford, Jauncey \& Murdoch 1970) to
fit the differential LFs with a simple
power-law model ($\frac{dN}{dL}\propto L^{-\beta}$). We note that the
exponent for a fit to the cumulative LF would be $\alpha=\beta-1$.
The best-fit slopes 
are shown in Table 1. Field 1 has the steepest slope of
$\alpha=1.7^{+0.34}_{-0.15}$ 
and normalization of 24 sources at $10^{36}$ \lum. The slopes of Field
2A (the section without the core region) and 3B (the part outside the
star-formation ring) are similar; Field 2A has a slope of
$0.9^{+0.16}_{-0.12}$ and
normalization of 24 sources at $10^{36}$ \lum, while Field 3 has a
slope of $1.1^{+0.20}_{-0.10}$ and
normalization of 27 sources at $10^{36}$ \lum.

\begin{inlinefigure}
\psfig{file=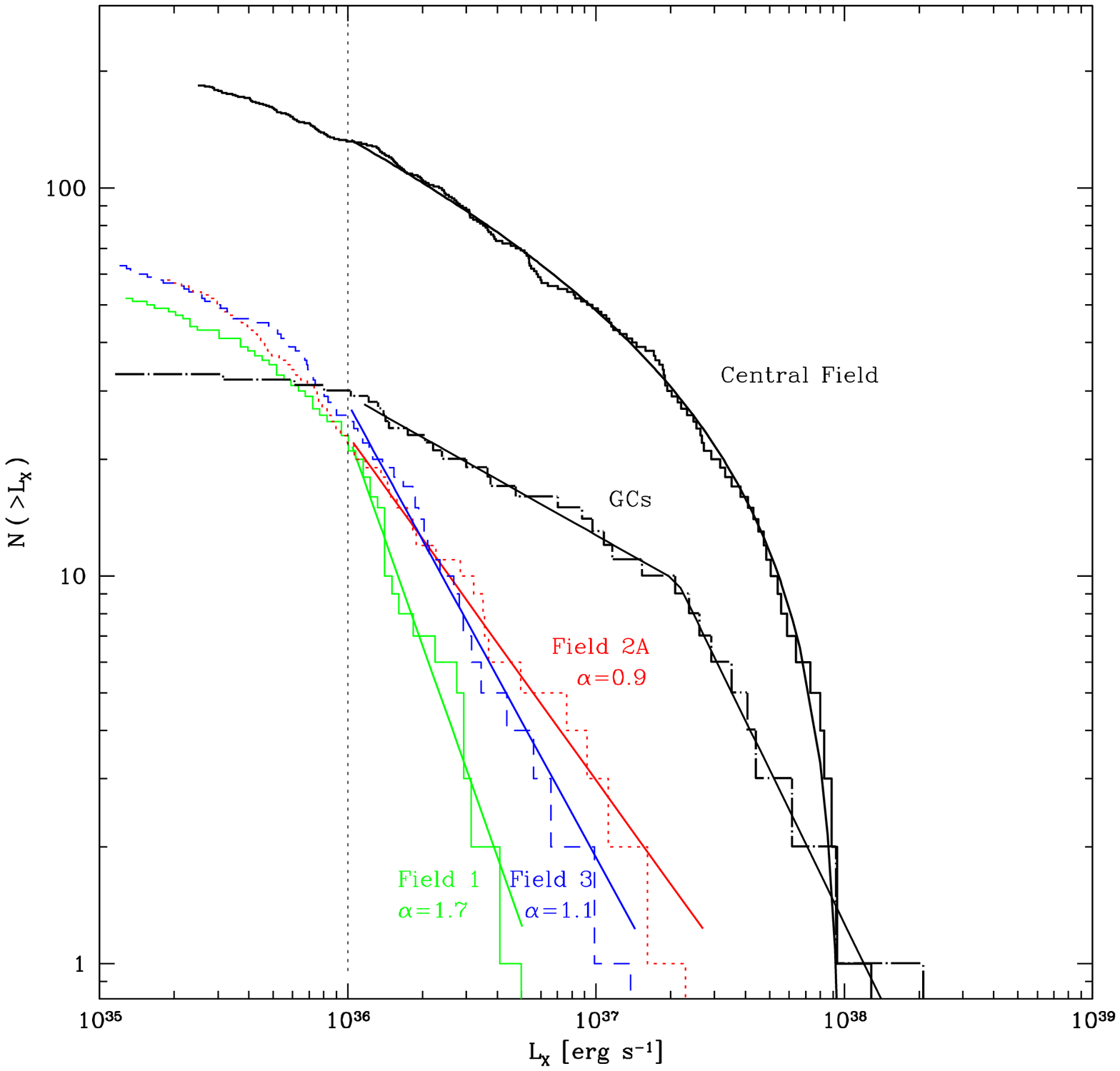,height=3.5in}
\caption{Cumulative luminosity functions and their best fit model for
Field 1 (green line), 2
(red line), 3 (blue line) and globular clusters (dotted-dash
line). The LF of
the bulge (solid black histogram) and its best-fit cutoff power-law model
(solid black curve) are shown for reference. The vertical
dotted line represents the completeness limit ($10^{36}$ \lum) of our
data.}
\end{inlinefigure}

\begin{table*}
\caption{Luminosity Functions}
{\centering
\begin{tabular}{lcccc}
\hline
\hline
 & \multicolumn{2}{c}{Power-law} &\multicolumn{2}{c}{Cutoff Power-law}\\
\cline{2-3} \cline{4-5}\\
&Slope ($\alpha$) & Break & Slope ($\alpha$)& Cutoff\\
&  & ($10^{37}$ \lum) &  & ($10^{37}$ \lum) \\
\hline
Field 1 & $1.7^{+0.34}_{-0.15}$&\\
Field 2A& $0.9^{+0.16}_{-0.12}$&\\
Field 3B & $1.1^{+0.20}_{-0.10}$&\\
r1\tablenotemark{a} & $0.88^{+0.32}_{-0.26}$, $0.73^{+0.15}_{-0.13}$
&$\sim 0.2$\\
r2\tablenotemark{b} & $0.58^{+0.11}_{-0.10}$,
$0.78^{+0.21}_{-0.17}$&$\sim 0.7$\\
r3\tablenotemark{c}& $0.55\pm0.06$, $1.93^{+0.54}_{-0.47}$ &
$\sim 2$\\
r1+r2+r3 & $0.50^{+0.06}_{-0.03}$,
$1.58^{+0.28}_{-0.25}$&$\sim 2$ & $0.25\pm0.07$ &
$9.83\pm0.40$ \\ 
GCs     & $0.3\pm0.18$, $1.2\pm0.08$& $\sim 2$\\
Integrated\tablenotemark{d} & $0.88\pm0.04$, $1.26\pm0.05$ &
$\sim 2$ & $0.49\pm0.07$ &
$10.60^{+0.90}_{-0.40}$  \\
\hline
\end{tabular}
\par
\medskip
\begin{minipage}{0.8\linewidth}
\footnotesize

NOTE.--- Uncertainties are 1-$\sigma$. The two values in the
power-law model are the slope below and above the luminosity break.\\

$^a$ Central $2'\times2'$.\\
$^b$ Central $8'\times8'$ excluding r1.\\
$^c$ Central $17'\times17'$ excluding r1+r2.\\
$^d$ Includes all sources from the central region
(r1+r2+r3; Kong et al. 2002) and the disk (Field 1, 2A, and 3).

\end{minipage}
\par
}
\end{table*}

Some of the sources may be background AGN. We estimated the
contribution of background objects by using 
number counts from the \chandra\ Deep Field survey (e.g., Brandt et
al. 2001; Giacconi et al. 2001). We found that at $\sim 10^{36}$ erg s$^{-1}$, about 10
sources in each field should be background objects. We checked this result
by counting the numbers of serendipitous point sources in each
of several fields from the ChaMP (\chandra\ Multiwavelength Project)
archives (P. Green, private communication); in each
of 5 ACIS-S observations of duration comparable to our exposure times,
the number of sources with count rates that would correspond
to M31 sources with $L_x > 10^{36}$ ergs s$^{-1}$ is $<
10$. Therefore, background AGN and foreground
stars have only a modest effect on the LF. By specifically subtracting
the effect of the 
background LF (e.g., Brandt et al. 2001; Giacconi et
al. 2001), we found that the uncertainty in the slope
due to background effects is within the 1$\sigma$
uncertainty limits for 
Field 2A and Field 3B. For Field 1, the slope steepens when we
include background effects, 
from 1.7 to 3.6, suggesting that the LF of Field 1 listed in Table 1
is very likely a lower limit.

\section{Interpretation}
We have two goals. (1) To relate the X-ray properties of
point sources in each group (Fields 1, 2, 3, the GCs, and the
central region) to properties (such as age) of the 
underlying stellar population.
(2) To
understand what the results imply for
X-ray observations of
distant galaxies that may be similar to M31.

We begin by counting the number of optically identified OB associations 
(Magnier et
al. 1993), supernova remnants (SNRs; d'Odorico et al. 1980; Braun \&
Walterbos 1993; Magnier et al. 1995), planetary nebulae (PN; Ford \&
Jacoby 1978; Ciardullo et al. 1989; Ciardullo, private communication ) and GCs (Battistini
et al. 1987; Magnier 1993; Barmby \& Huchra 2001) in each of the three
fields.
Although these catalogs may not be complete and certainly suffer
from a variety of selection effects, 
they do provide a gross indication
of the stellar populations inhabiting each field.
Table 2 summarizes the results.
 
We now proposed to study the X-ray properties of each field and
relate them to the stellar populations that inhabit the field.
Because each field was the subject of identical X-ray observations,
comparisons of X-ray properties across fields are well-defined.
The situation is not as straightforward for
optical observations.

\subsection{Optical Observations}

All of our $3$ fields were surveyed for GCs (Battistini
et al. 1987; Magnier 1993; Barmby \& Huchra 2001) and PN (Ford \&
Jacoby 1978; Ciardullo et al. 1989; Ciardullo, private communication) 
The entire galaxy was also surveyed by
ISO far-infrared (175$\mu$m) observations
(Schmidtobreick, Haas, \& Lemke 2000). 

Data is also available for optically-identified OB associations
(Magnier et
al. 1993) and for 
supernova remnants (SNRs; d'Odorico et al. 1980; Braun \&
Walterbos 1993; Magnier et al. 1995).
Field 2, which is closest to
the galaxy center, was fully covered by each of these surveys,
but only 
roughly half of
Field 1, and \lax \,20\% of Field 3, as described below,
were covered.

\subsection{The Fields}
 
\noindent {\bf Field 1:\ } 
Field 1 is in the far southwestern end of the 
disk, $\sim 70'$ or 16 kpc from the galactic center. 
Since all fields received identical X-ray coverage,
we can say that the spatial density
of X-ray sources in Field 1 is  lower than in the other fields,
the slope of the LF is steeper,  
and the maximum point-source X-ray luminosity is smaller. In fact,
the population is dominated by faint X-ray sources ($<
10^{37}$ \lum).
The only optical counterparts we found are 4 
possible foreground stars (Di\,Stefano et al. 2002b). 
Two far-infrared knots are located in the
north-east corner of the field (Schmidtobreick, Haas, \&
Lemke 2000), which might hint at the presence of
a young population in this portion of the field.
This is consistent with the presence of 
a clump of 10 SNRs near the far-infrared knots.
These $10$ SNRs were discovered by Magnier et al.\ (1995) in
a survey that covered approximately half of Field 1. No
additional Field 1 SNRs were found in that survey, in spite
of the fact the area of Field 1 that was surveyed was $\sim 10$
times larger than the region defined by the SNR clump. In addition, 
no OB associations were found in Field 1 (Magnier 1993),
although only roughly half of the field was surveyed.
One additional SNR was discovered in Field 1 (d'Odorico 1980) 
in a photographic
survey.   
The combination of these data 
indicate that only a small portion
of the northern half of Field 1
contains a young stellar population.
This implies that roughly half
of the Field 1 X-ray sources lie in regions not associated with
young stellar populations.  
Apart from the absence of far-infrared knots in the southern portion
of Field 1, we have no clear indication about the presence or absence of
SNRs or OB associations.
We therefore refer to the integrated optical light. According to
Walterbos \& Kennicutt (1988), the total integrated
$B$ magnitude is roughly $4$ magnitudes dimmer
at the location of Field 1 than near the galaxy center,
and the values of $U-B$, $B-V$, and $B-R$, show that this region
is clearly redder than the central region. This trend
is supported by other investigations (see Hodge \& Kennicutt 1982,
and references in Hodge 1992). On the basis of this continuum light,
it would be surprising if the southern half of Field 1
housed a large young population of stars. We note further,
that, for the total numbers of SNRs and OB associations  
in Field 1 to be comparable to that in Fields 2 or 3A,
their local density in the southern half of Field 1
would have to be larger than their density even in
Field 2.           

\begin{table*}
\caption{Source statistics in the fields of view}
{\centering
\begin{tabular}{ccccccc}
\hline
\hline
& Field 1 & \multicolumn{2}{c}{Field 2} && \multicolumn{2}{c}{Field 3}\\
\cline{3-4} \cline{6-7}\\
&         &A & B (nucleus) &&A (young)& B (old) \\
\hline
$N_{X-ray}$ (including GCs)& 53 & 68 & 31\tablenotemark{c} && 16 & 49\\
$N_{X-ray~GC}$ & 0 & 11 & 2\tablenotemark{c} & &0&1\\
$N^{opt}_{GC}$ (fraction\tablenotemark{a}\,) & 16 (25.4\%)& 145 (42\%) &57 (27\%)&& 30 (27\%)&
45 (71\%)\\
$N^{opt}_{PN}$ (fraction) & 36 (57.1\%)& 113 (33\%) & 151 (72\%)& &  18
(16\%)&18 (29\%)\\
$N^{opt}_{SNR}$ (fraction)& 11 (17.5\%)& 65 (19\%) & 0 (0\%) && 48 (42\%)
&0 (0\%) \\
$N^{opt}_{OB}$ (fraction) & 0  (0\%) & 19 (6\%) & 2 (1\%)&& 17 (15\%) & 0
(0\%) \\\
Average $L_x$\tablenotemark{b} & 1.14 & 2.35 & 13&& 1.15 & 1.82 \\ 
Maximum $L_x$ & 5 & 26 & 60 & &3.3 & 420\\
\hline
\end{tabular}
\par
\medskip
\begin{minipage}{0.8\linewidth}
\footnotesize

NOTE.--- For comparison, there are 12 GC X-ray sources in all
fields, while
11 of them are in Field 2A. The average and maximum 0.3--7 keV luminosity of GCs is
$2.8\times10^{37}$\lum\ and $4\times10^{38}$\lum, respectively.
$N_{X-ray}$ is the total number of X-ray sources. $N^{opt}$ is the number 
of particular optical sources. 

$^a$ Fraction is defined as number of
this type of optical object divided by the total number of optical sources
(GCs+PN+SNRs+OB associations).\\
$^b$ 0.3--7 keV luminosity (excluding GCs) in unit of
$10^{36}$\lum.\\
$^c$ This number is an underestimate, because the spatial
resolution in this region is significantly degraded (see text).

\end{minipage}
\par
}
\end{table*}

\noindent {\bf Field 2:\ } 
Field 2 has the highest spatial density of
X-ray point sources. 
The average and maximum X-ray source luminosities
are also highest in Field 2. 
The slope of its LF
is significantly smaller than that  of Field 1,
and is
similar to slopes measured in starburst galaxies and in the
star-formation regions of spirals (e.g., Pence et al. 2001; Tennant et
al. 2001; Kilgard et al. 2002; Soria \& Kong 2002).
Field 2 is close to the galaxy center, where
the density of GCs is highest; 11 X-ray sources in
Field 2A (the portion near the aim-point) have been identified with
GCs (see also Di\,Stefano et al. 2002a). 
One source in Field 2 appears to be coincident with an SNR. 
Far-infrared
observations (Haas et al. 1998; Schmidtobreick et al. 2000)
have hinted at star
formation activities near the bulge. In particular, there are 9
bright far-infrared knots (Schmidtobreick et al. 2000) that are
coincident with the X-ray sources in Field 2.
The density of cataloged
optical sources is highest; more than 80 optical sources are
SNRs or OB associations, indicating a significant presence of young
stars. It therefore seems likely 
that the X-ray properties of at least a portion of the 
population in Field 2 is 
characteristic of young stellar populations.

\noindent{\bf Field 3:\ } 
In terms of its distance from the nucleus, 
the X-ray properties of its point sources,
and the numbers and types of optical sources,
Field 3 lies between Fields 1 and 2.   
The slope of the LF of Field 3 
is determined almost entirely by sources outside the
star-forming ring (Field 3B).
In fact, 
the inclusion or exclusion of the sources from Field 3A 
(the section near the star 
forming ring) does not significantly  affect the fit,
since only 5
sources (above $10^{36}$\lum) belong to Field 3A. 

Field 3A was well-covered by the surveys for SNR and OB 
associations, and    
visual
examination of the optical fields (Magnier et al. 1995) demonstrates
 that
these objects are highly concentrated in the star-formation ring. 
The region adjacent to the ring was also covered by the surveys. 
We therefore know that there is a sharp spatial cut-off in the
distribution of SNRs and OB associations, with no indication
of them away from the ring. There is, therefore, no
sign that the regions not surveyed for SNRs and OB
associations are rich with these objects.
This conclusion is consistent with the distribution
of far-infrared knots: there are $3$ such knots, and,
although the entire field was searched, 
all $3$ are located
along the star-forming ring. 
The total integrated
$B$ magnitude is roughly $3$ magnitudes dimmer
at the location of Field 1 than near the galaxy center
(Walterbos \& Kennicutt 1988),
and, as for Field 1,
 the values of $U-B$, $B-V$, and $B-R$, show that the region
is redder than the central region
(see also Hodge \& Kennicutt 1982)
It therefore seems unlikely that Field 3B houses a large
young population. 
 
It is perhaps  puzzling that the slope of Field 3's LF
 is essentially the
same as for Field 2 (within the uncertainty limits), but is
less steep than 1. 

\noindent{\bf GCs:\ }
The LF of GCs differs from that of all three disk fields and the central
region. We consider all GC X-ray sources (34 in total) that have been
observed by \chandra, including both disk (Di\,Stefano et al. 2002a,2002b)
and bulge (Kong et al. 2002a). The power-law
model has a break at $\sim
2\times10^{37}$\lum, similar to the central region but the slopes (below
and above the break) of the LF
are flatter than that of the central region (Kong et al. 2002a). It is
worth noting that
the luminosity break is also near the luminosity of the brightest GC in our 
Galaxy (see
Di\,Stefano et al. 2002a). The average and maximum luminosity of GCs
is also significantly higher than that of Fields 1, 2 and 3 (see Table 2
and Figure 2).

\noindent{\bf Integrated LF:\ } 
To compare M31 with more distant galaxies, 
it is useful to study the composite M31 LF. This consists of data from
both central region (Kong et al. 2002a) and disk regions (see 
Figure 3). As in the central region of M31 (e.g., Primini, Forman, \&
Jones
1993; Shirey et al. 2001; Kong et al. 2002a; Kaaret 2002), there is a luminosity break at
$\sim 2\times 10^{37}$ \lum. This break might
represent an aging X-ray binaries population
(Wu 2001; Kilgard et al. 2002; Kaaret 2002).
We fit the differential LF with two power-law models with a
break at $\sim 2\times10^{37}$\lum\ and the results are summarized in Table 1. The shape 
of the integrated LF differs from that of the central region, as
Fields 1, 2A and 3 have less luminous
sources (\lax \,$2\times10^{37}$\lum) and have relatively steep
slopes for the LF. The slope of the integrated LF below $\sim
2\times10^{37}$\lum\ is therefore steeper than that of the bulge region (see
Table 1). The slope of the integrated LF above the break is consistent with that of GCs, 
suggesting that luminous X-ray sources of M31 are dominated by GCs. 

\begin{inlinefigure}
\psfig{file=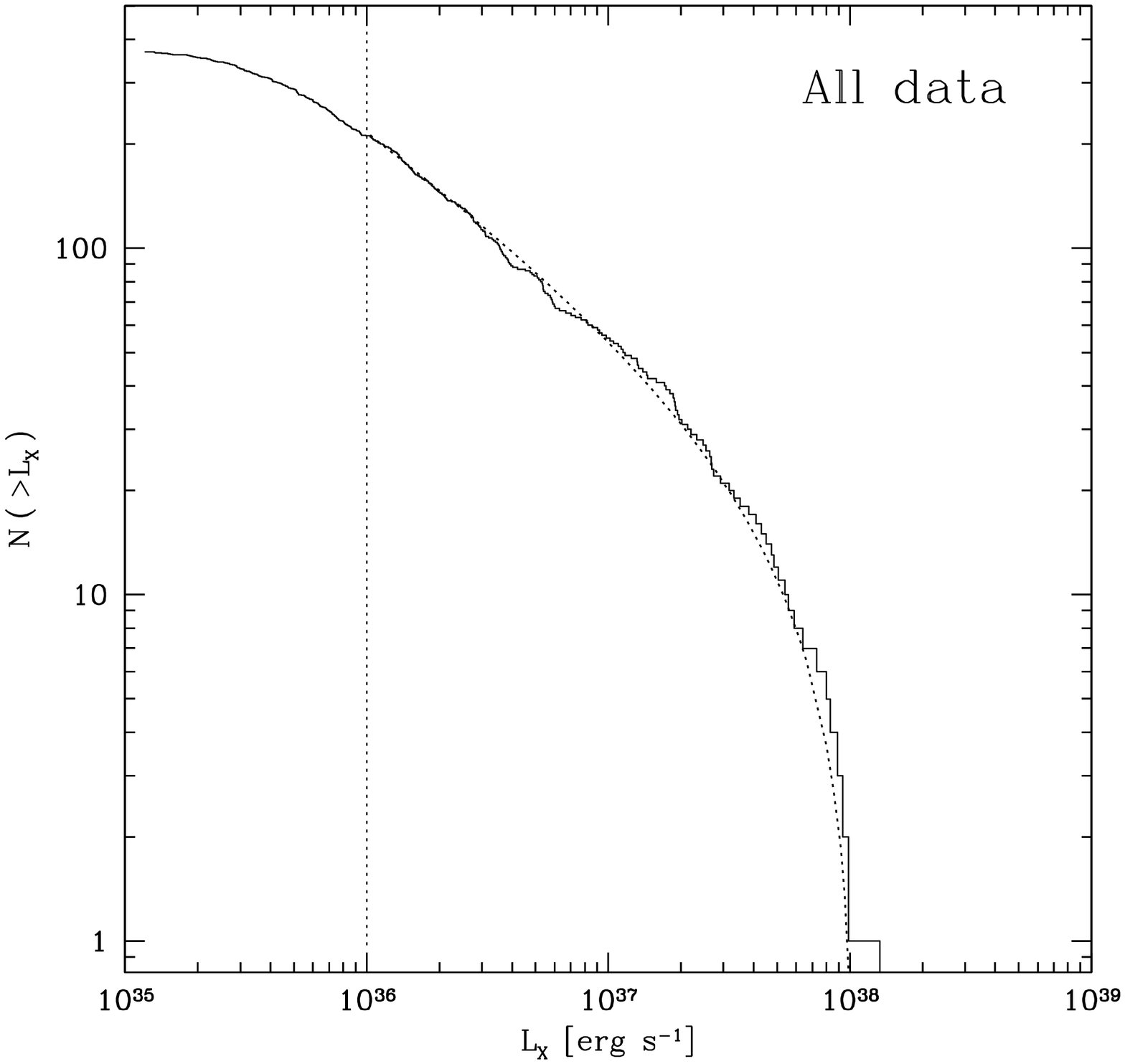,height=3.5in}
\caption{Integrated (combining the central $17'\times17'$ region,
Field 1, 2 and 3) cumulative luminosity function of M31. The
vertical dotted line represents the completeness limit ($10^{36}$
\lum) of our data. The dotted line show the cutoff power-law fit with
cutoff luminosity
at $10^{38}$\lum\ for the data.}
\end{inlinefigure}

We also fit the differential LF with a cutoff power-law model 
($\frac{dN}{dL}\propto L^{-\beta}e^{-L/L_{cut}}$; Grimm et al. 2002
and Trudolyubov et al. 2002). The best-fitting slope ($\alpha=\beta-1$)
and
cutoff luminosity, $L_{cut}$, are 0.49 and $1.06\times10^{38}$ \lum,
respectively. Compared to the central $17'\times17'$ region, the
slope is steeper while the cutoff luminosity is roughly the same (see
Table 1). It is worth noting that the cutoff power-law fit to the central
$17'\times17'$ region is in good agreement with the central $15'$
observed with \xmm\ (Trudolyubov et al. 2002).
The best fit parameters of the integrated LF are consistent with the LF
of low-mass X-ray binaries (LMXBs) in our own
Galaxy (Grimm et al. 2002), which could indicate that LMXBs are
important components of M31's X-ray population.

\section{Conclusions}

\subsection{Connections Between Optical 
and X-Ray Properties}

With the large number of X-ray sources in each of several different
stellar environments, it is appropriate to look for patterns that
may relate the LF of each region to its stellar properties.
Developing such a relationship could be useful in the  study of
more distant galaxies in which the X-ray sources and/or the stellar
environments are not as well resolved as in M31.

Judging by the numbers and types of optical sources in Fields 1, 3, 
and 2, it seems as if there is a progression from an old population (1),
to another old population which adjoins a region
of star formation (3), to the very mixed  population near the galaxy
center, which includes
a significant sub-population of young stars (2).  The X-ray properties
also seem to exhibit a progression. The slope of the LF
is steepest in Field 1, and less steep in Fields 2 and 3. The density of
X-ray  sources increases
from Field 1, to Field 3, and is highest in Field 2. 
The value of the maximum source luminosity increases as well.

In spite of these apparent trends, it is wise to 
be cautious, as there are also some apparent discrepancies 
when we make comparisons with complementary data sets.
We first consider ACIS-I observations of the galaxy center, which have 
also been used to construct LFs (Kong et al. 2002a). The two
inner regions, r1 and r2 encompass $8' \times 8'$ around the galaxy center;
together they include 92 point sources. The LFs in these 2 regions are
similar; the composite has slope $0.80\pm0.13$ above the luminosity
break (see Kong et al. 2002a).      
That this is obviously comparable to the slope of Field 2A's LF, 
since 2A and (r1 + r2) are close to each
other. The puzzle is that r3, which covers the central
$17' \times 17'$ (excluding r1+r2), and includes $112$ point sources,
has a markedly different slope,  $1.93^{+0.54}_{-0.47}$, much closer
to the slope of Field 1. 

A second relevant comparison can be made with \xmm\
observations of
the northern disk (North1 + North2; Figure 1; Trudolyubov et al. 2002).
The LF also has a slope ($\alpha=1.3\pm0.2$)
intermediate between that of Field 1 and Field 3B. Yet, judging by
the location of
OB associations and SNRs (see Figure 1), the stellar population would
appear likely to have much in common with Field 3A and perhaps Field 2.
Indeed, by comparing with our Galaxy (Grimm et
al. 2002), Trudolyubov
et al. (2002) concluded that the northern disk of M31 is dominated by faint
and young high-mass X-ray binaries (HMXBs). 
It is worth noting that Grimm et
al. (2002) derived a slope of $0.63\pm0.13$ for the cumulative LF of
HMXBs in our Galaxy, which is significantly flatter than that of
the \xmm\ northern disk fields. In constrast, the LF of Field 2A is
closer to the Galactic HMXB population (but note that there are
uncertainties due to the differences in energy coverage of \chandra,
\xmm\ and {\it RXTE}/ASM and other
instrumental effects). 

These examples make it clear that the slope, taken by itself,
is not a good
indication of the age of the underlying stellar
population, and hence of the age and character of the X-ray sources. 
Yet, while the slopes of LFs for dissimilar stellar populations may be
the same, other qualities (e.g., average and maximum luminosities, and
source densities)
differ, and these combined with information about X-ray colors
(Prestwich et al. 2002) may break the degeneracy.

\subsection{GCs} 
The data on M31's GCs demonstrate two things. First,  
there is an important difference between a subset of the X-ray GCs
in M31 and the X-ray GCs
in the Galaxy. In fact all of the M31 GC sources above the high-luminosity 
break are more luminous than any Galactic GC X-ray source; i.e., this
part of the M31 GC luminosity function has no Galactic analog.
It has been argued on the basis of binary evolution, that 
these high-$L_X$ sources may be a signal that a subset of M31's
GCs may be younger than Galactic GCs (Di Stefano et al. 2002a).
The possible future evolution of the LF toward lower luminosities
may also support this interpretation.
Second, if the shape of the LF is a reliable guide,
there is an important difference between M31's X-ray  
sources in GCs and the X-ray sources in both the central field and the
disk. If a significant portion of non-GC sources are objects that were
once ejected from GCs, it will be necessary to determine how the
GC LF can evolve into either the LF of sources in the
central field (where most of the GCs reside) or into the
LFs found for disk sources.  

\subsection{The Integrated Luminosity Function of M31}

The integrated LF makes it clear that observers in distant galaxies would
not find M31 to be a very impressive X-ray galaxy. If  observers
in the Virgo cluster were to observe M31 with the equivalent of
a 40 ksec ACIS-S observations, only a handful (4) of sources (those
with $L_x$ \gax \,$10^{38}$ \lum) would be visible at
any given time. Half of these would be near the center of the galaxy;
the remaining sources would be in GCs.
Field 1 would have no detectable sources; Field 3A would be, at best,
sparsely populated. This result is consistent with earlier surveys
(Supper et al. 1997,2001) and
with \xmm\ data (Shirey et al. 2001; Trudolyubov et
al. 2002). Although our own Galaxy may house a somewhat
larger number of sources with $L_x$ \lax \,$10^{38}$\lum (Grimm et
al. 2002), 
the basic
result holds for our Galaxy as well, with the difference that it is likely
that none of the detectable sources would be in GCs. This result makes 
it clear that the galaxy we live in and its nearest neighbor are very
different from the distant galaxies on which many current X-ray studies
concentrate. 

\begin{acknowledgements}
We are grateful to Paul Green and the ChaMP collaboration (see
http://hea-www.harvard.edu/CHAMP/) for providing blank field data,
and Phil Kaaret for discussions. This work
was supported in part by NASA under GO1-2091X, NAG5-10889 and NAG5-10705. 
\end{acknowledgements}

\end{document}